\documentclass[%
 reprint,
 superscriptaddress,
 amsmath,amssymb,
 aps,
prl,
floatfix,
]{revtex4-2}

\usepackage{graphicx}
\usepackage{dcolumn}
\usepackage{bm}
\usepackage{hyperref}
\usepackage{xcolor}

\begin{document}

\preprint{APS/123-QED}

\title{Observation of {biased} random-flux-induced topological phase transition in gyromagnetic photonic crystals}

\author{Hai-Xiao Wang}%
\email{wanghaixiao@nbu.edu.cn}
\affiliation{School of Physical Science and Technology, Ningbo University, Ningbo 315211, China}%

\author{Chongyang Li}
\affiliation{State Key Laboratory of Optical Fiber and Cable Manufacture Technology, Department of Electronic and Electrical Engineering, Guangdong Key Laboratory of Integrated Optoelectronics Intellisense, Southern University of Science and Technology,Shenzhen 518055, China}

\author{Xianmu Wu}
\affiliation{School of Physical Science and Technology, Ningbo University, Ningbo 315211, China}%
\affiliation{School of Physical Science and Technology, Guangxi Normal University, Guilin 541004, China}

\author{Ziyao Wang}
\affiliation{State Key Laboratory of Optical Fiber and Cable Manufacture Technology, Department of Electronic and Electrical Engineering, Guangdong Key Laboratory of Integrated Optoelectronics Intellisense, Southern University of Science and Technology,Shenzhen 518055, China}
\author{Yongmei Wang}
\affiliation{School of Physical Science and Technology, Guangxi Normal University, Guilin 541004, China}
\author{Junhui Hu}
\affiliation{School of Physical Science and Technology, Guangxi Normal University, Guilin 541004, China}

\author{Shiwei Tang}%
\affiliation{School of Physical Science and Technology, Ningbo University, Ningbo 315211, China}%

\author{Zhen Gao}
\email{gaoz@sustech.edu.cn}
\affiliation{State Key Laboratory of Optical Fiber and Cable Manufacture Technology, Department of Electronic and Electrical Engineering, Guangdong Key Laboratory of Integrated Optoelectronics Intellisense, Southern University of Science and Technology,Shenzhen 518055, China}

\date{\today}

\begin{abstract}
	The interplay between disorder and topological states has attracted growing interest. While previous studies have primarily addressed the effects of geometric or potential randomness, the exploration of topological phase transitions driven by random-flux {remains} experimentally elusive. Here, we report the first experimental realization of topological phase transitions driven by biased random-flux in gyromagnetic photonic crystals. By stochastically orienting the magnetization of constituent gyromagnetic rods, we implement a disordered Haldane model in which the sign of the next-nearest-neighbor hopping phases is randomly distributed. We demonstrate that the bulk band gap closes when the densities of positive and negative flux are balanced—restoring time-reversal symmetry in a statistical sense—and reopens when a net positive or negative flux is introduced. Microwave near-field measurements directly visualize the reversal of chiral edge states, confirming a transition between distinct topological phases. Our results establish a unique disorder-driven mechanism for realizing topological phase transitions and deepen our understanding of the interplay between disorder and topological phases in bosonic systems.
	 
	\end{abstract}
	
	\maketitle
	
	
{\it Introduction}---Topological physics has fundamentally reshaped our understanding of phase transitions since the discovery of the quantum Hall effect~\cite{toporev1,toporev2,toporev3,toporev4,topophotonrev,topophononrev1,topophononrev2,topoelectricrev}. A representative example is the Haldane model~\cite{haldane}, which was initially proposed as a paradigmatic model to demonstrate that quantized Hall conductance can emerge from the breaking of time-reversal symmetry in a lattice with zero net magnetic flux. Subsequent experimental realizations of the Haldane model have spanned various platforms, including {cold atom}~\cite{haldane_electron}, photonic~\cite{haldane_photon1,haldane_photon2,haldane_photon3}, acoustic~\cite{haldane_phonon1}, and mechanical~\cite{haldane_phonon2} systems, demonstrating that topological properties are typically governed by the phase of complex hopping terms. While strong disorder in electronic systems traditionally induces Anderson localization, recent studies on topological Anderson insulators~\cite{Anderson_1,Anderson_2,Anderson_3,Anderson_4} have revealed that disorder can counterintuitively facilitate the emergence of nontrivial topological phases~\cite{Anderson_photon_1,Anderson_photon_2,Anderson_photon_3,Anderson_photon_4,Anderson_photon_5,Anderson_photon_6,Anderson_phonon_1,Anderson_phonon_2,Anderson_phonon_3,Anderson_circuit_1,Anderson_circuit_2,Anderson_quasi_1,Anderson_quasi_2,Anderson_nonherm_1,Anderson_nonherm_2,Anderson_nonherm_3}. 

However, the vast majority of research on disorder-induced topology has focused on the geometric or potential randomness. A more profound yet less explored frontier is the impact of random-flux, where disorder resides in the hopping phase rather than the onsite potential. In the context of the Haldane model, this corresponds to random-flux disorder wherein the local magnetic flux, namely, the sign of the next-nearest-neighbor (NNN) hopping phase, fluctuates spatially. Recent studies on topological photonic alloys~\cite{Anderson_alloy_1,Anderson_alloy_2,Anderson_alloy_3} verified a disorder-induced trivial-to-nontrivial topological phase transition, where a random local breakdown of time-reversal symmetry give rise to a Chern insulating phase from a trivial background. A fundamental yet unexplored question is whether a topological phase transition can be driven between two distinct nontrivial phases by the competition between two opposite-sign fluxes. On the other hand, random-flux-induced topological phase transition was theoretically demonstrated in two-dimensional Su-Schrieffer-Heeger model~\cite{random_flux_1}, anisotropic Wilson-Dirac model~\cite{random_flux_2} and disordered Haldane model~\cite{random_flux_3}, respectively. To date, however, a random-flux-induced topological phase transition has yet to be observed experimentally, owing to the challenge of precisely controlling random-flux. 

\begin{figure*}[htbp]
	\centering\includegraphics[width=6.8in]{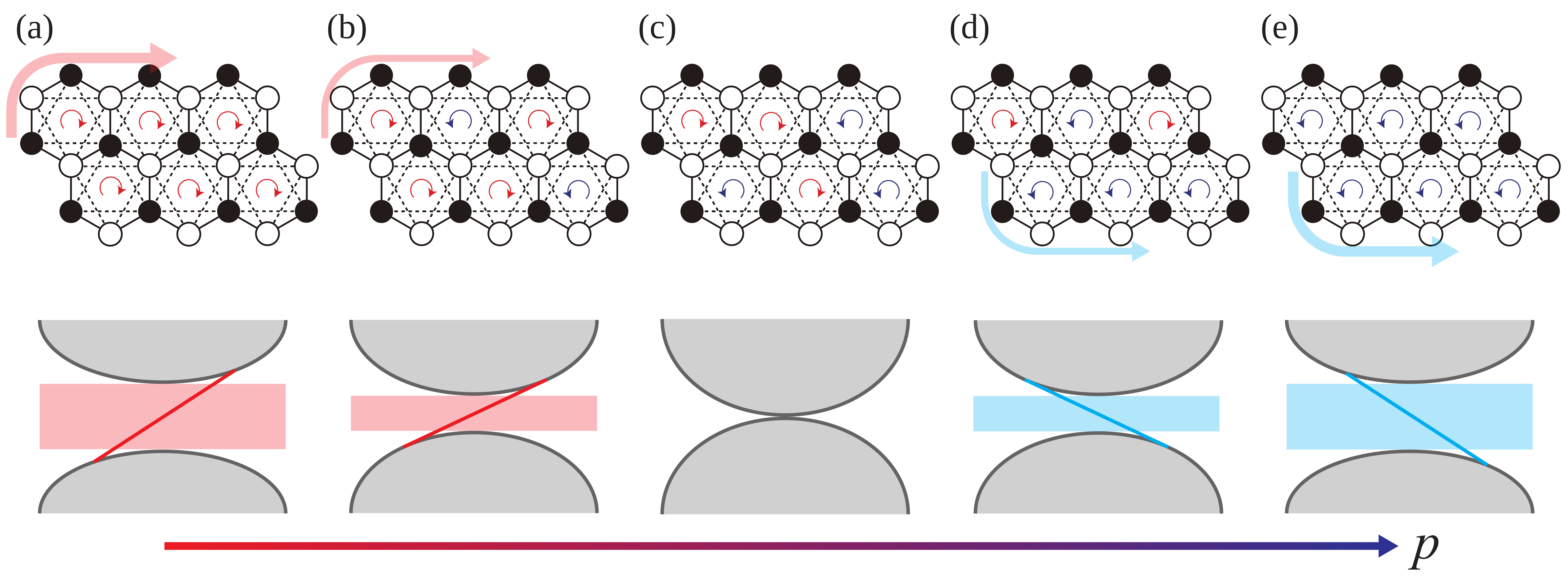}
	\caption{Illustration of {biased} random-flux-induced topological phase transition. (a, e) A clean Haldane model with all NNN hopping phase are (a) positive, and (e) negative. (b-d) Disordered Haldane model with negative-flux concentration (b) $p<0.5$, (c) $p=0.5$, and (d) $p>0.5$ by continuously {tuning} $p$ from 0 to 1. The positive- and negative-flux are indicated by clockwise and counterclockwise arrows, respectively. Lower panel: the illustration of the projected band structure and chiral edge states, where the red and blue area refer to the band gap with  $C=1$ and $C=-1$, respectively.}
	\label{Fig_1}
\end{figure*} 

To fill this gap, we experimentally demonstrate a topological phase transition driven by biased random-flux in a gyromagnetic photonic crystal (GPC). Our system consists of a stochastic mixture of yttrium iron garnet (YIG) rods with opposite magnetization directions but identical magnitudes, which maps onto a disordered Haldane model with randomly assigned signs for the NNN hopping phases. Taking the inversion symmetric Haldane model as an illustration, the positive (negative) NNN hopping phase $\phi$ ($-\phi$) indicated by the clockwise (counterclockwise) arrow opens a spectral gap $E_g=2|m_0|$ at the Dirac points, where $m_0$ is proportional to $\sin \phi$ and represents the Dirac mass [see Figs.~\ref{Fig_1}(a) and \ref{Fig_1}(e)], supporting a gapless edge states whose chirality is determined by $sgn(m_0)$ at the spectral gap. When the NNN phases are stochastically assigned signs of $\pm \phi$ with a negative-flux concentration $p$ [see Figs.~\ref{Fig_1}(b) and \ref{Fig_1}(d))], the effective mass can be determined via the statistical average of the local mass terms, i.e., $\bar{m} =(1-p)\cdot m_0+p\cdot (-m_0)=(1-2p)m_0$. Hence, the system becomes gapped when the net flux is positive or negative biased, supporting a gapless edge state whose chirality is determined by $sgn(\bar{m})$. The relationship between the biased random-flux and the spectral gaps also predicts a critical gap closing point at $p=0.5$, where the equal population of opposite fluxes restores time-reversal symmetry in a statistical sense [see Fig.~\ref{Fig_1}(c) and details in Supplemental Materials (SM)]. Experimentally, the biased random-flux in disordered Haldane model could be effectively controlled by the concentration ratio of the negatively magnetized YIG rods. Using microwave near-field measurements, we provide direct evidence of this transition through the visualization of chiral edge state reversal. Our work establishes a robust platform for exploring the interplay between random gauge fields and topological order, offering a unique mechanism for manipulating bosonic states through flux-sign disorder.

\begin{figure}[htbp]
	\centering\includegraphics[width=3.4in]{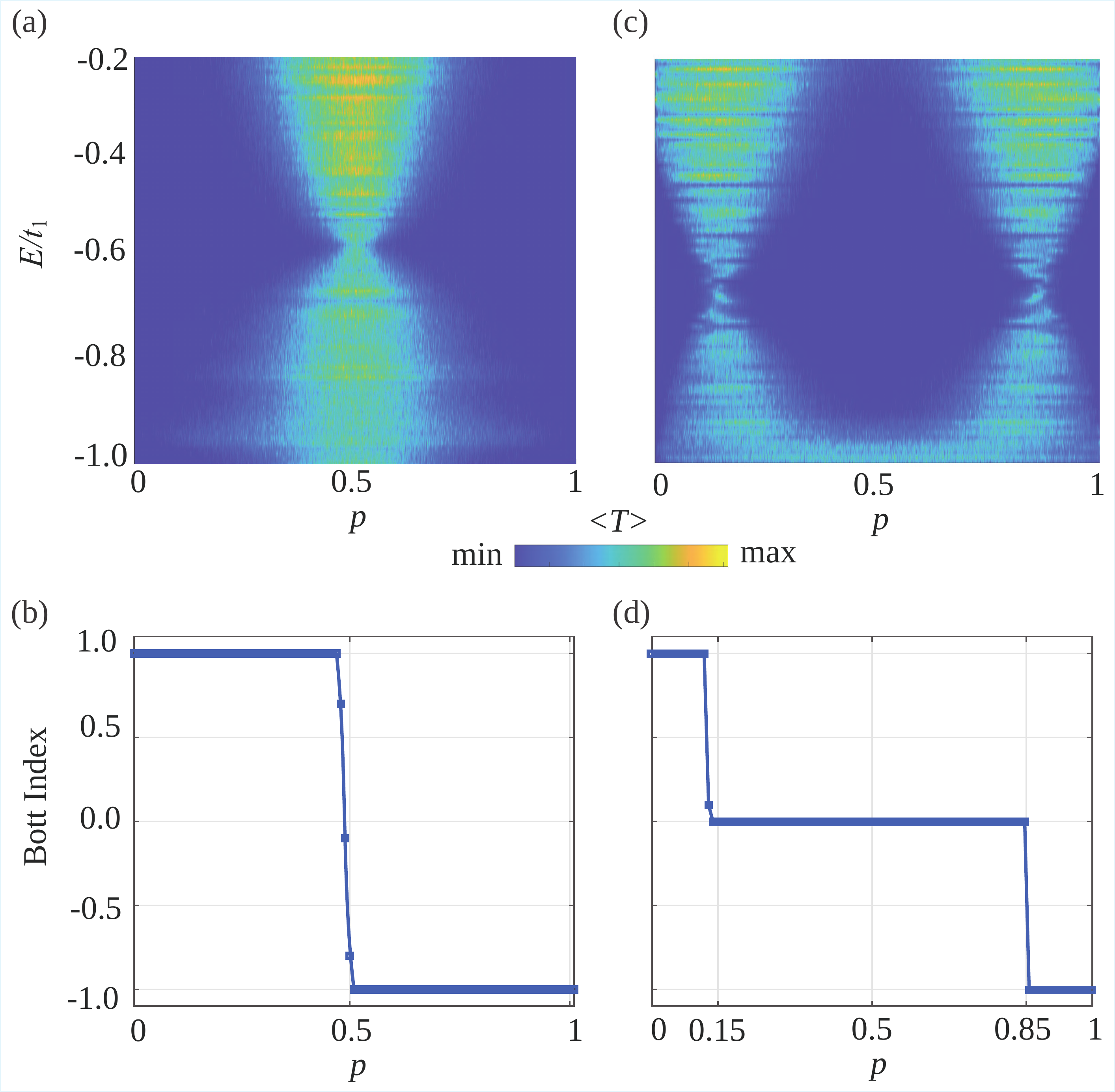}
	\caption{(a,c) The bulk transmission spectra and (b,d) the Bott index versus the negative-flux concentration $p$ with (a,b) $M=0$, and (c,d) $M=0.9$. Note that the size of the calculated supercell is $15a_0 \times 15 \sqrt3 a_0$ with lattice constant $a_0=1$, and both the bulk transmission spectra and the Bott index are derived by averaging across 10 configurations. Other parameters are as follows: $t_1=1,\ t_2=\frac{1}{3},\phi=\frac{\pi}{4}$.}
	\label{Fig_2}
\end{figure}

{\it Theoretical analysis}---To verify the {biased} random-flux-induced topological phase transition, we revisit the Haldane model, whose tight-binding Hamiltonian is given by
\begin{equation}
	H=\sum_{i}M\epsilon_i c_i^\dagger c_i+ t_1\sum_{\langle i,j\rangle}c_i^{\dagger}c_j+t_2\sum_{\langle\langle i,j\rangle\rangle}e^{iv_{ij}\phi}c_i^{\dagger}c_j+h.c., 
\end{equation}
{where $c_i^\dagger $ and $c_i$ are the creation and annihilation operators on the honeycomb lattice sites}, respectively. $M$ denotes the on-site energy, while $t_1$ and $t_2$ represent to the nearest and next-nearest neighbor hopping amplitudes, respectively. $\epsilon_i=1 (-1)$ for the {different} sublattice sites. The term $v_{ij}=\pm 1$ accounts for the alternating sign of the magnetic phase of the NNN couplings that breaks the time-reversal symmetry when $\phi\neq n\pi$. We then construct a disordered Haldane model by randomly alternating the sign of $\phi$, where the ratio of the negative flux is defined as the concentration $p$. Next, we examine the transport properties of the systems with varying concentration by calculating the two-terminal conductance using the recursive Green’s function method~\cite{Green_Function_1,Green_Function_2}. At this stage, we assume that inversion symmetry is preserved by setting $M=0$, and two semi-finite normal metal leads are attached to the end of the disordered Haldane model (see details in SM). It is seen from Fig.~\ref{Fig_2}(a) that the bulk band gap indicated by the transmission gap undergoes a closing and reopening process. To characterize the topological properties of the disordered Haldane model with varying negative-flux concentration $p$, we further calculate {the} Bott index below the energy of the critical phase transition point (see details in SM)~\cite{Bott_index_1,Bott_index_2}. The result in Fig.~\ref{Fig_2}(b) shows that the Bott index of disordered Haldane model gives $+1 (-1)$ when $p<0.5 (p>0.5)$ and suffers a nearly sudden jump around $p=0.5$. The consistency between the transmission spectra and the Bott index versus negative-flux concentration $p$ identifies the {biased} random-flux-induced phases transition. From the perspective of Weyl physics, the critical phase transition shares similar physics with a quadratic Weyl point~\cite{Weyl_physics_1,Weyl_physics_2,Weyl_physics_3,Weyl_physics_4}, which separates two quantum Hall phases with opposite Chern numbers ($C=1$ and $C=-1$) and hosts a topological charge of $-2$. Furthermore, we consider the modified Haldane model without inversion symmetry by setting $M=0.9$. The transmission result in Fig.~\ref{Fig_2}(c) shows that the transmission gap closes around at approximately $p=0.15$ and $p=0.85$. Accordingly, the Bott index of disordered Haldane model in Fig.~\ref{Fig_2}(c) gives $+1 ,0,-1$ when $p<0.15$,$0.15<p<0.85$, and $p>0.85$, respectively, indicating the existence of two distinct critical phase transition points. The first phase transition separates the disordered Haldane model from a quantum Hall phase with $C=1$ to a trivial insulator phase with $C=0$, while the second transition converts the system into another quantum Hall phase with $C=-1$. From the perspective of Weyl physics, the breaking of inversion symmetry results in the splitting of a quadratic Weyl point, generating two Weyl points carrying the same topological charge. 
 
\begin{figure*}[htbp]
	\centering\includegraphics[width=6.8in]{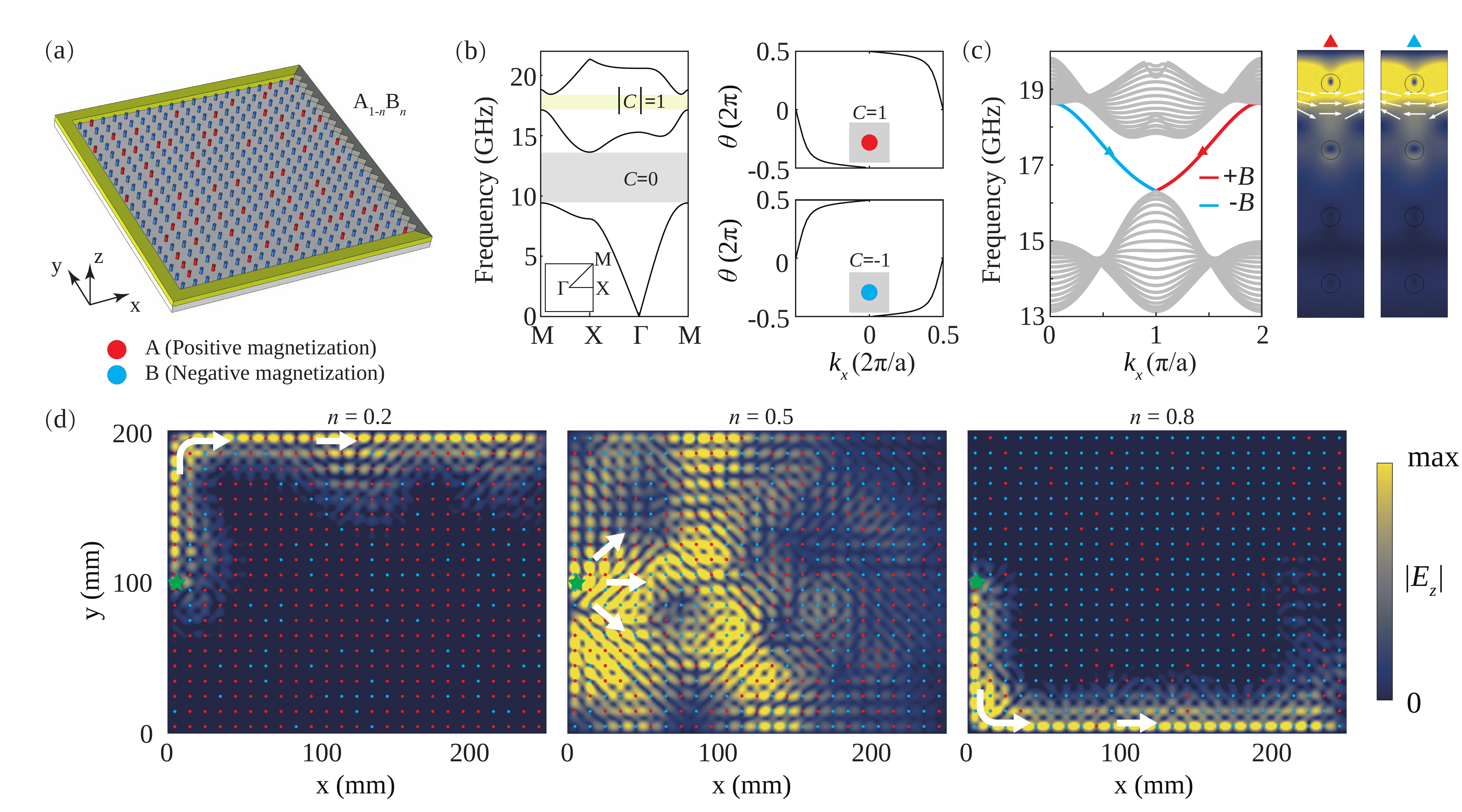}
	\caption{(a) Schematic of the disordered GPCs $A_{1-n}B_n$. (b) Left panel: photonic band structure of $A-$ and $B-$ type GPCs. Right panel: the Wilson loop of $A-$ (upper) and $B-$ (lower) type GPCs, giving the gap Chern number with 1 and -1, respectively. (c) Left panel: The projected band structure (gray lines) and the upper edge states (red and blue lines) of $A-$ and $B-$ type GPCs under perfect electric conductor condition. Right panel: the Poynting vector distributions of upper chiral states in $A-$ and $B-$ type GPCs. (d) Simulated electric field distribution at $f=17.44$ GHz for disordered GPCs $A_{0.8}B_{0.2}$ (left panel), $A_{0.5}B_{0.5}$ (middle panel), and $A_{0.2}B_{0.8}$ (right panel). Green stars mark the position of line sources.}
	\label{Fig_3}
\end{figure*}

To demonstrate this {biased} random-flux-induced phase transition, we design a square lattice GPCs $A_{1-n}B_n$ with a size of $25a \times 20a$ [see Fig.~\ref{Fig_3}(a), where $a=11.5$mm is the lattice constant], comprising a mixture of positively ($A$) and negatively ($B$) magnetized YIG rods. The subscript $n$ represents the concentration of negatively magnetized rods, i.e., $n=N_B/(N_A+N_B)$, with $N_A$ and $N_B$ representing the number of $A-$ and $B-$type rods, respectively. Throughout this work, we only focus on the transverse magnetic modes.    

In the clean limit ($n=0$ and $n=1$), the system reduces to a standard photonic Chern insulator featuring unidirectionally propagating edge states. Figure ~\ref{Fig_3}(b) presents the band structures and the corresponding Chern numbers in the bandgaps for pure $A-$ and $B-$type GPCs. Obviously, they share identical band structures but possess opposite Chern numbers. As depicted in the right panel of Fig.~\ref{Fig_3}(b), the Wannier centers of the second band gap for pure $A-$ and $B-$type GPCs exhibit the same winding number but opposite winding directions, indicating the gap Chern numbers of two systems are opposite. Accordingly, we implement the eigen calculation of a ribbon strip supercell clad with perfect electric conductors under opposite external magnetic fields. As shown in Fig.~\ref{Fig_3}(c), a gapless edge state with a positive group velocity (red line) emerges in the bulk band gap when the external magnetic field is applied along $+z$ direction. Flipping the magnetic field reverse the group velocity of the gapless edge state (blue line). Accordingly, the Poynting vector distributions of the upper chiral states in $A-$ and $B-$ types GPCs are opposite at $f=17.51$GHz [see right panel of Fig.~\ref{Fig_3}(c)], further confirming that they have opposite Chern numbers.

Next, we consider three representative disordered GPCs with $n=0.2$, $n=0.5$, and $n=0.8$, respectively (see the configuration details in SM). Full-wave stimulated field patterns under the excitation of a point source at $f=17.44$ GHz are shown in Fig.~\ref{Fig_3}(d). Note that the all disordered GPCs are enclosed by metal cladding on three sides and an absorbing material on the right side [also see Fig.~\ref{Fig_3}(a)]. As shown in the left and right panels of Fig.~\ref{Fig_3}(d), the numerical results indicate the presence of chiral edge states bound between the disorder GPCs and the metal cladding when the concentration $n=0.2$ and $n=0.8$. These states propagate unidirectionally and can robustly wrap around corners, confirming that both the disordered GPCs $A_{0.8}B_{0.2}$ and $A_{0.2}B_{0.8}$ possess topological band gaps. Importantly, their propagation directions are opposite, verifying that they carry opposite Chern number. In contrast, the electric field pattern for $A_{0.5}B_{0.5}$ only suggests an excited bulk states. According to the bulk-edge correspondence, this {confirms} that a phase transition occurs as the concentration varies from 0 to 1. {Furthermore, we characterize the topological properties of the band gaps by calculating the reflection phases of the supercell (see details in SMs)}. The results show that at a frequency of 17.44GHz, the disordered GPCs $A_{0.2}B_{0.8}$ and $A_{0.8}B_{0.2}$ share the same phase winding number but opposite winding direction when the incident phase angle varied by $2\pi$, indicating they are of opposite Chern numbers. 

\begin{figure*}[htbp]
	\centering\includegraphics[width=6.8in]{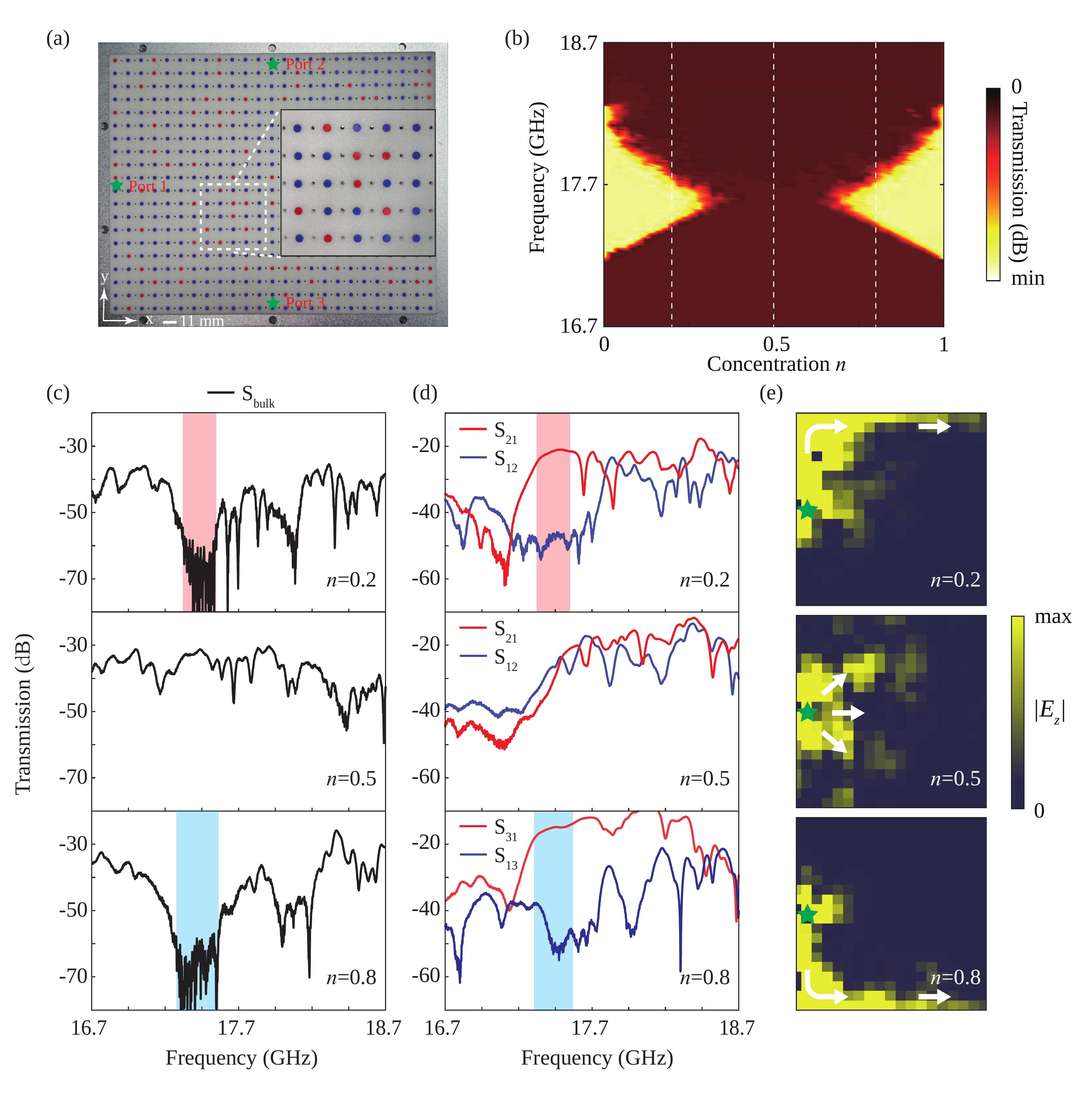}
	\caption{(a) Photograph of the fabricated sample. The perforated air foams are used to fix the magnet/YIG/magnet (red and blue colors) rods. Inset: Zoom-in of the perforated air foams. (b) Simulated phase diagram of the disordered GPCs versus concentration $n$ {using 20 configurations}. Three dashed lines refer to $n=0.2$,$n=0.5$, and $n=0.8$, respectively. (c) Measured bulk transmission spectra. Red and blue regions indicate the measured bulk band gaps (defined by $S_{bulk}<-60$dB), which are 17.33-17.53GHz and 17.27-17.56GHz for $n=0.2$ and 0.8, respectively. No band gap is observed for $n=0.5$. (d) Measured edge transmission spectra. Red and blue regions refer to the frequency range of the nonreciprocal edge propagation (defined by $|S_{31}-S_{13}|>30$dB). No unidirectional edge transportation is observed for $n=0.5$. (e) The measured electric field intensity distribution excited by a point source at a frequency of 17.66 GHz.}
	\label{Fig_4}
\end{figure*}	

{\it Experimental observation of {biased} random-flux-induced topological phase transition}---To verify our theoretical predictions, we then fabricated an experimental sample and carried out microwave near-field measurements. As shown in Fig.~\ref{Fig_4}(a), the sample has a size of $25a \times 20a$. Parallel aluminum plates are perforated with air holes to allow the excitation excite and probe the bulk and edge states and the perforated air foams are used to fix the magnet/YIG/magnet rods (indicated by red and blue circles). In the experiment, both positively and negatively magnetized YIG rods are individually biased using permanent magnets embedded beneath them. We display the calculated phase diagram of the disordered GPCs in Fig.~\ref{Fig_4}(b). {Note that bulk transmission are obtained by averaging over 20 configurations at each frequency.} The simulated transmission indicates that the bulk gap vanishes in the concentration range of $0.35<n<0.65$, which differs slightly from our prediction. This deviation is primarily attributed to the finite size effects of the selected supercell. Figure ~\ref{Fig_4}(c) presents the measured bulk transmission spectra for $n=0.2,0.5$, and 0.8, respectively. It is observed that bulk transmissions experience sudden drops ranging from 17.33-17.53GHz and 17.27-17.56GHz when $n=0.2$ and $n=0.8$, respectively, indicating the existence of two bulk band gaps (defined as frequency ranges where $S_{bulk}<-60$dB, see red and blue rectangular areas). In contrast, the negligible difference between forward and backward bulk transmission for $n=0.5$ indicates the closing of the bulk band gap. Besides, we also implement the edge transmission measurement $S_{ij}$ by placing a point source at the left side of the cladding edge (port 1), and the detected probe at upper (port 2) and lower (port 3) metal claddings. For $n=0.2$, it is seen from Fig.~\ref{Fig_4}(d) that the clockwise edge transmission spectra $S_{21}$ is (around 30dB) higher than that with anticlockwise edge transmission $S_{12}$, giving evidence of nonreciprocity of the chiral edge states. For $n=0.8$, similar nonreciprocal transmission spectra, i.e., $|S_{31}-S_{13}|>30$dB, are observed in the frequency gap. {Note that a slight frequency shift can be observed between the rise of the $S_{31}$ signal and the drop of the $S_{13}$ signal around the lower boundary of the bulk gap (the blue region in Fig. 4(d)), which is primarily attributed to the finite-size effect of our experimental sample.}  Nevertheless, the chiral edge state propagates counterclockwise, which is opposite to the propagation direction for $n=0.2$, demonstrating the gap Chern numbers of the two cases are opposite. For $n=0.5$, the edge transmission spectra exhibit reciprocity feature due to the absence of a bulk band gap. The measured bulk and edge transmission characteristics are explicit evidence of a {biased} random-flux driven transition between two distinct topological nontrivial phases. 

To directly visualize the chiral edge propagation, we mapped the field intensity point by point inside the sample. As shown in Fig.~\ref{Fig_4}(e), when a point source at a frequency of 17.66 GHz is excited at the left boundary of the sample, the emitted fields for $n=0.2$ are strongly confined to the sample edges, and can propagate clockwise with negligible backscattering. Conversely, it is seen from the emitted field that edge states propagate counterclockwise without significant loss for $n=0.8$. The reversal of the propagation direction confirms the sign flip of the gap Chern number. As a comparison, the field distribution for $n=0.5$ reveals an extended mode at a frequency of 17.66 GHz, identifying it is a bulk state.  

{\it Conclusion}---In conclusion, we have experimentally demonstrated a topological phase transition induced by {biased} random-flux in disordered GPCs consisting of a stochastic mixture of positively and negatively magnetized rods. The disordered GPCs map directly onto a disordered Haldane model with biased random-flux. When the densities of positive and negative flux are balanced, the system restores time-reversal symmetry in a statistical sense accompanied by the closing of the bulk band gap. Using near-field measurements, we directly observed that the chiral edge states reverse their propagation direction as the net flux crosses a critical point, verifying a topological phase transition driven purely by random disorder. Our work provides an alternative avenue for realizing topological phase transitions and deepen our understanding of the interplay between disorder and topological phases in bosonic systems.     

{\it Data availability}---The data that support the findings of this articles are not publicly available. The data are available from the authors upon reasonable request.

{\it Funding Statement}---H.-X. W. thank the support from National Natural Science Foundation of China (Grants No. 12474432), Natural Science Foundation of Guangxi Province (Grant No. 2023GXNSFAA026048), and the start-up funding of Ningbo University. Z. G. acknowledges funding from the National Key R\&D Program of China (Grant No. 2025YFA1412300), National Natural Science Foundation of China (Grants No. 62361166627 and No. 62375118), Guangdong Basic and Applied Basic Research Foundation (Grant No. 2024A1515012770), Shenzhen Science and Technology Innovation Commission (Grant No. 202308073000209), and Highlevel Special Funds (Grant No. G03034K004).

{\it Author Contributions}---H.X.W. and Z. Wang contributed equally to this letter.

{\it Conflicts of Interest Disclosure} --- The authors declare no conflicts of interest.


\begin{thebibliography}{99}
	
	\bibitem{toporev1}{M. Z. Hasan, C. L. Kane, {\it Colloquium}: Topological insulators, Rev. Mod. Phys. {\bf 82}, 3045 (2010). \url{https://www.doi.org/10.1103/RevModPhys.82.3045}}
	
	\bibitem{toporev2}{X.-L. Qi, S.-C. Zhang, Topological insulators and superconductors, Rev. Mod. Phys. {\bf 83}, 1057 (2011).\url{https://www.doi.org/10.1103/RevModPhys.83.1057}}
	
	\bibitem{toporev3}{N. P. Armitage, E. J. Mele, A. Vishwanath, Weyl and Dirac semimetals in three-dimensional solids, Rev. Mod. Phys. {\bf 90}, 015001 (2018). \url{https://www.doi.org/10.1103/RevModPhys.90.015001}}
	
	\bibitem{toporev4}{B. Xie, H.-X. Wang, X. Zhang, P. Zhan, J.-H. Jiang, M. Lu, Y. Chen, Higher-order band topology, Nat. Rev. Phys. {\bf 3}, 520 (2021).\url{https://www.doi.org/10.1038/s42254-021-00323-4}}
		
	\bibitem{topophotonrev}{T. Ozawa, H. M. Price, A. Amo, N. Goldman, M. Hafezi, L. Lu, M. C. Rechtsman, D. Schuster, J. Simon, O. Zilberberg, I. Carusotto, Topological photonics, Rev. Mod. Phys. {\bf 91}, 015006 (2019). \url{https://www.doi.org/10.1103/RevModPhys.91.015006}}
	
	\bibitem{topophononrev1}{H. Xue, Y. Yang, B. Zhang, Topological acoustics, Nat. Rev. Mater. {\bf 7}, 974 (2022).\url{https://www.doi.org/10.1038/s41578-022-00465-6}}
	
	\bibitem{topophononrev2} {W. Zhu, W. Deng, Y. Liu, J. Lu, H.-X. Wang, Z.-K. Lin, X. Huang, J.-H. Jiang, Z. Liu, Topological phononic metamaterials, Rep. Prog. Phys. {\bf 86}, 106501 (2023).\url{https://www.doi.org/10.1088/1361-6633/aceeee}}
	
	\bibitem{topoelectricrev}{H. Yang, L. Song, Y. Cao, P. Yan, Circuit realization of topological physics, Phys. Rep. {\bf 1093}, 1 (2024).\url{https://www.doi.org/10.1016/j.physrep.2024.09.007}}
	
	\bibitem{haldane}{F. D. M. Haldane, Model for a Quantum Hall Effect without Landau Levels: Condensed-Matter Realization of the "Parity Anomaly", Phys. Rev. Lett. {\bf 61}, 2015 (1988).\url{https://www.doi.org/10.1103/PhysRevLett.61.2015}}
	
	\bibitem{haldane_electron}{G. Jotzu, M. Messer, R. Desbuquois, M. Lebrat, T. Uehlinger, D. Greif, T. Esslinger, Experimental realization of the topological Haldane model with ultracold fermions, Nature {\bf 515}, 237 (2014).\url{https://www.doi.org/10.1038/nature13915}}
	
	\bibitem{haldane_photon1}{Z. Wang, Y. Chong, J. D. Joannopoulos, M. Soljačić, Observation of unidirectional backscattering-immune topological electromagnetic states, Nature {\bf 461}, 772 (2009)\url{https://www.doi.org/10.1038/nature08293}}.
	
	\bibitem{haldane_photon2}{G.-G. Liu, P. Zhou, Y. Yang, H. Xue, X. Ren, X. Lin, H. Sun, L. Bi, Y. Chong, B. Zhang, Observation of an unpaired photonic Dirac point, Nat. Commun. {\bf 11}, 1873 (2020).\url{https://www.doi.org/10.1038/s41467-020-15801-z}}
	
	\bibitem{haldane_photon3}{G.-G. Liu, Z. Gao, Q. Wang, X. Xi, Y.-H. Hu, M. Wang, C. Liu, X. Lin, L. Deng, S. A. Yang, P. Zhou, Y. Yang, Y. Chong, B. Zhang, Topological Chern vectors in three-dimensional photonic crystals, Nature {\bf 609}, 925 (2022).\url{https://www.doi.org/10.1038/s41586-022-05077-2}}
	
	\bibitem{haldane_phonon1}{Y. Ding, Y. Peng, Y. Zhu, X. Fan, J. Yang, B. Liang, X. Zhu, X. Wan, J. Cheng, Experimental Demonstration of Acoustic Chern Insulators, Phys. Rev. Lett. {\bf 122}, 014302 (2019).\url{https://www.doi.org/10.1103/PhysRevLett.122.014302}}
	
	\bibitem{haldane_phonon2}{L. M. Nash, D. Kleckner, A. Read, V. Vitelli, A. M. Turner, W. T. M. Irvine, Topological mechanics of gyroscopic metamaterials, Proc. Natl. Acad. Sci. U.S.A. {\bf 112}, 14495 (2015).\url{https://www.doi.org/10.1073/pnas.1507413112}}
	
	\bibitem{Anderson_1}{J. Li, R.-L. Chu, J. K. Jain, S.-Q. Shen, Topological Anderson Insulator, Phys. Rev. Lett. {\bf 102}, 136806 (2009).\url{https://www.doi.org/10.1103/PhysRevLett.102.136806}}
	
	\bibitem{Anderson_2}{C. W. Groth, M. Wimmer, A. R. Akhmerov, J. Tworzydło, and C. W. J. Beenakker, Theory of the Topological Anderson Insulator, Phys. Rev. Lett. {\bf 103}, 196805 (2009).\url{https://www.doi.org/10.1103/PhysRevLett.103.196805}}	
	
	\bibitem{Anderson_3}{H. Jiang, L. Wang, Q. Sun, X. C. Xie, Numerical study of the topological Anderson insulator in HgTe/CdTe quantum wells, Phys. Rev. B {\bf 80}, 165316 (2009).\url{https://www.doi.org/10.1103/PhysRevB.80.165316}}
	
	\bibitem{Anderson_4}{A. Neehus, F. Pollmann, and J. Knolle, Genuine Topological Anderson Insulator from Impurity Induced Chirality Reversal, Phys. Rev. Lett. {\bf 135}, 126604 (2025).\url{https://www.doi.org/10.1103/10.1103/7p8y-2mp6}}
		
	\bibitem{Anderson_photon_1}{C. Liu, W. Gao, B. Yang, S. Zhang, Disorder-Induced Topological State Transition in Photonic Metamaterials, Phys. Rev. Lett. {\bf 119}, 183901 (2017).\url{https://www.doi.org/10.1103/PhysRevLett.119.183901}}
	
	\bibitem{Anderson_photon_2}{S. Stützer, Y. Plotnik, Y. Lumer, P. Titum, N. H. Lindner, M. Segev, M. C. Rechtsman, and A. Szameit, Photonic topological Anderson insulators, Nature {\bf 560}, 461 (2018).\url{https://www.doi.org/10.1038/s41586-018-0418-2}}
	
	\bibitem{Anderson_photon_3}{G.-G. Liu, Y. Yang, X. Ren, H. Xue, X. Lin, Y.-H. Hu, H. Sun, B. Peng, P. Zhou, Y. Chong, B. Zhang, Topological Anderson Insulator in Disordered Photonic Crystals, Phys. Rev. Lett. {\bf 125}, 133603 (2020).\url{https://www.doi.org/10.1103/PhysRevLett.125.133603}}
	
	\bibitem{Anderson_photon_4}{X. Cui, R.-Y. Zhang, Z.-Q. Zhang, C. T. Chan, Photonic Z2 Topological Anderson Insulators, Phys. Rev. Lett. {\bf 129}, 043902 (2022).\url{https://www.doi.org/10.1103/PhysRevLett.129.043902}}
	
	\bibitem{Anderson_photon_5}{Y. Yang, X. Qian, L. Shi, X. Shen, and Z. H. Hang, Unidirectional transport in amorphous topological photonic crystals, Sci. China Phys. Mech. Astron. 66, 274212 (2023).\url{https://www.doi.org/10.1007/s11433-023-2093-9}}
	
	\bibitem{Anderson_photon_6}{X.-D. Chen, Z.-X. Gao, X. Cui, H.-C. Mo, W.-J. Chen, R.-Y. Zhang, C. T. Chan, J.-W. Dong, Realization of Time-Reversal Invariant Photonic Topological Anderson Insulators, Phys. Rev. Lett. {\bf 133}, 133802 (2024).\url{https://www.doi.org/10.1103/PhysRevLett.133.133802}}	
	
	\bibitem{Anderson_phonon_1}{F. Zangeneh-Nejad, R. Fleury, Topological optomechanically induced transparency, Opt. Lett. {\bf 45}, 5966 (2020).\url{https://www.doi.org/10.1364/OL.410002}}
	
	\bibitem{Anderson_phonon_2}{F. Zangeneh‐Nejad, R. Fleury, Disorder Induced Signal Filtering with Topological Metamaterials, Adv. Mater. {\bf 32}, 2001034 (2020).\url{https://www.doi.org/10.1002/adma.202001034}}
	
	\bibitem{Anderson_phonon_3}{H. Liu, B. Xie, H. Wang, W. Liu, Z, Li, H. Cheng, J. Tian, Z. Liu, S. Chen, Acoustic spin-Chern topological Anderson insulators, Phys. Rev. B {\bf 108}, L161410 (2025). \url{https://www.doi.org/10.1103/PhysRevB.108.L161410}}
	
	\bibitem{Anderson_circuit_1}{Z.-Q. Zhang, B.-L. Wu, J. Song, H. Jiang, Topological Anderson insulator in electric circuits, Phys. Rev. B {\bf 100}, 184202 (2019).\url{https://www.doi.org/10.1103/PhysRevB.100.184202}}
	
	\bibitem{Anderson_circuit_2}{W. Zhang, D. Zou, Q. Pei, W. He, J. Bao, H. Sun, X. Zhang, Experimental Observation of Higher-Order Topological Anderson Insulators, Phys. Rev. Lett. {\bf 126}, 146802 (2021).\url{https://www.doi.org/10.1103/PhysRevLett.126.146802}}
		
	\bibitem{Anderson_quasi_1}{T. Peng, C.-B. Hua, R. Chen, D.-H. Xu, B. Zhou, Topological Anderson insulators in an Ammann-Beenker quasicrystal and a snub-square crystal, Phys. Rev. B {\bf 103}, 085307 (2021).\url{https://www.doi.org/10.1103/PhysRevB.103.085307}}
	
	\bibitem{Anderson_quasi_2}{X. Cheng, T. Qu, L. Xiao, S. Jia, J. Chen, L. Zhang, Topological Anderson amorphous insulator, Phys. Rev. B {\bf 108}, L081110 (2023).\url{https://www.doi.org/10.1103/PhysRevB.108.L081110}}	
	
	\bibitem{Anderson_nonherm_1}{H. Jiang, L.-J. Lang, C. Yang, S.-L. Zhu, S. Chen, Interplay of non-Hermitian skin effects and Anderson localization in nonreciprocal quasiperiodic lattices, Phys. Rev. B {\bf 100}, 054301 (2019).\url{https://www.doi.org/10.1103/PhysRevB.100.054301}}
	
	\bibitem{Anderson_nonherm_2}{Z. Gu, H. Gao, H. Xue, D. Wang, J. Guo, Z. Su, B. Zhang, Observation of an acoustic non-Hermitian topological Anderson insulator, J. Zhu, Sci. China Phys. Mech. Astron. {\bf 66}, 294311 (2023).\url{https://www.doi.org/10.1007/s11433-023-2159-4}}
	
	\bibitem{Anderson_nonherm_3}{B.-B. Wang, Z. Cheng, H.-Y. Zou, Y. Ge, K.-Q. Zhao, Q.-R. Si, S.-Q. Yuan, H.-X. Sun, H. Xue, B. Zhang, Observation of disorder-induced boundary localization, Proc. Natl. Acad. Sci. U.S.A. {\bf 122}, e2422154122 (2025).\url{https://www.doi.org/10.1073/pnas.2422154122}}	
	
	\bibitem{Anderson_alloy_1}{T. Qu, M. Wang, X. Cheng, X. Cui, R.-Y. Zhang, Z.-Q. Zhang, L. Zhang, J. Chen, C. T. Chan, Topological Photonic Alloy, Phys. Rev. Lett. {\bf 132}, 223802 (2024). \url{https://www.doi.org/10.1103/PhysRevLett.132.223802}}
	
	\bibitem{Anderson_alloy_2}{Z. Wang, X. Xi, Z. Gao, Topological photonic quasicrystal alloy, Appl. Phys. Lett. {\bf 125}, 122201 (2024).\url{https://www.doi.org/10.1063/5.0232244}}
	
	\bibitem{Anderson_alloy_3}{B. Huang, Z. Wang, Z. Gao, Amorphous topological photonic alloys, Phys. Rev. B {\bf 112}, 14204 (2025).\url{https://www.doi.org/10.1103/gw3w-5h94}}
	
	\bibitem{random_flux_1}{C.-A. Li, S.-B. Zhang, J. C. Budich, and B. Trauzettel, Transition from metal to higher-order topological insulator driven by random-flux, Phys. Rev. B {\bf 106}, L081410 (2022).\url{https://www.doi.org/10.1103/PhysRevB.106.L081410}}
	
	\bibitem{random_flux_2}{C.-A. Li, B. Fu, J. Li, B. Trauzettel, Random-flux-induced transition sequence between weak and strong topological phases with anisotropic localization properties, Phys. Rev. B {\bf 111}, 214207 (2025).\url{https://www.doi.org/10.1103/zjyw-ln2n}}
	
	\bibitem{random_flux_3}{J. Wang and W.-M. Liu, random-flux manipulating topological phase transitions in Chern insulators, Chin. Phys. B {\bf 34}, 067301 (2025).\url{https://www.doi.org/10.1088/1674-1056/adc36a}}
	
	\bibitem{Green_Function_1}{S. Datta, Electronic Transport in Mesoscopic Systems, Cambridge Univ. Press, Cambridge, 2007.\url{https://www.cambridge.org/core/product/identifier/9780511805776/type/book}}
	
	\bibitem{Green_Function_2}{G. Metalidis and P. Bruno, Green’s function technique for studying electron flow in two-dimensional mesoscopic samples, Phys. Rev. B {\bf 72}, 235304 (2005).\url{https://www.doi.org/10.1103/PhysRevB.72.235304}}
	
	\bibitem{Bott_index_1}{Z.-R. Liu , C.-B Hua, T. Peng, and B Zhou, Chern insulator in a hyperbolic lattice, Phys. Rev. B {\bf 105}, 245301 (2022).\url{https://www.doi.org/10.1103/PhysRevB.105.245301}}
	
	\bibitem{Bott_index_2}{D. Toniolo, On the Bott index of unitary matrices on a finite torus, Lett. Math. Phys. {\bf 112}, 126 (2022).\url{https://www.doi.org/10.1103/10.1007/s11005-022-01602-6}}
	
	\bibitem{Weyl_physics_1}{T. Zhang, Z. Song, A. Alexandradinata, H. Weng, C. Fang, L. Lu, and Z. Fang, Double-Weyl Phonons in Transition-Metal Monosilicides, Phys. Rev. Lett. {\bf 120}, 016401 (2018).\url{https://www.doi.org/10.1103/PhysRevLett.120.016401}}
	
	\bibitem{Weyl_physics_2}{H. He, C. Qiu, X. Cai, M. Xiao, M. Ke, F. Zhang, Z. Liu, Observation of quadratic Weyl points and double-helicoid arcs, Nat. Commun. {\bf 11}, 1820 (2020).\url{https://www.doi.org/10.1038/s41467-020-15825-5}}
	
	\bibitem{Weyl_physics_3}{C. Jörg, S. Vaidya, J. Noh, A. Cerjan, S. Augustine, G. Von Freymann, and M. C. Rechtsman, Observation of Quadratic (Charge‐2) Weyl Point Splitting in Near‐Infrared Photonic Crystals, Laser Photon. Rev. {\bf 16}, 2100452 (2022).\url{https://www.doi.org/10.1002/lpor.202100452}}
	
	\bibitem{Weyl_physics_4}{Z.-M. Yu, Z. Zhang, G.-B. Liu, W. Wu, X.-P. Li, R.-W. Zhang, S. A. Yang, Y. Yao, Encyclopedia of emergent particles in three-dimensional crystals, Sci. Bull. {\bf 67}, 375 (2022).\url{https://www.doi.org/10.1016/j.scib.2021.10.023}}
	
	\end{thebibliography}
\end{document}